\documentclass{article}

\usepackage{arxiv}

\usepackage{amssymb}
\usepackage{amsmath}
\usepackage{amsfonts}
\usepackage{xcolor}
\usepackage{url}
\usepackage{subfig}
\usepackage{soul}
\usepackage{pgfplots}
\usepackage{layouts}

\newcommand{\comment}[1]{}

\usepackage{listings}
\definecolor{codegreen}{rgb}{0,0.6,0}
\definecolor{codegray}{rgb}{0.5,0.5,0.5}
\definecolor{codepurple}{rgb}{0.58,0,0.60}
\definecolor{backcolour}{rgb}{0.95,0.95,0.95}

\lstdefinestyle{mystyle}{
    backgroundcolor=\color{backcolour},   
    commentstyle=\color{codegreen},
    keywordstyle=\color{magenta},
    numberstyle=\tiny\color{codegray},
    stringstyle=\color{codepurple},
    basicstyle=\ttfamily\footnotesize,
    frame = single,
    breakatwhitespace=false,         
    breaklines=true,                 
    captionpos=b,                    
    keepspaces=true,                 
    numbers=right,                    
    numbersep=5pt,                  
    showspaces=false,                
    showstringspaces=false,
    showtabs=false,                  
    tabsize=4
}
\usepackage{lineno}

\usepackage{float}
\restylefloat{table}

\title{j-Wave: An open-source differentiable wave simulator}

\author{
  Antonio Stanziola$^*$ \\
  Dept. of Medical Physics and Biomedical Engineering \\
  University College London \\
  London WC1E 6BT, UK\\
  \\
   \And
  Simon R. Arridge \\
  Department of Computer Science \\
  University College of London \\
  London WC1E 6BT, UK \\ 
  \\
   \And
  Ben T. Cox \\
  Dept. of Medical Physics and Biomedical Engineering \\
  University College London \\
  London WC1E 6BT, UK\\
  \\
  \And
  Bradley E. Treeby \\
  Dept. of Medical Physics and Biomedical Engineering \\
  University College London \\
  London WC1E 6BT, UK\\
  \\
}

\date{}

\begin{document}

\maketitle

\begin{abstract}
We present an open-source differentiable acoustic simulator, j-Wave, which can solve time-varying and time-harmonic acoustic problems. It supports automatic differentiation, which is a program transformation technique that has many applications, especially in machine learning and scientific computing. j-Wave is composed of modular components that can be easily customized and reused. At the same time, it is compatible with some of the most popular machine learning libraries, such as JAX and TensorFlow. The accuracy of the simulation results for known configurations is evaluated against the widely used k-Wave toolbox and a cohort of acoustic simulation software. j-Wave is available from \url{https://github.com/ucl-bug/jwave}.
\end{abstract}

\keywords{
differentiable simulator \and acoustics \and machine learning \and gpu acceleration \and wave equation \and Helmholtz equation \and jax
}



\comment{
\section*{Required Metadata}

\section*{Current code version}

\begin{table}[H]
\begin{tabular}{|l|p{6.5cm}|p{6.5cm}|}
\hline
\textbf{Nr.} & \textbf{Code metadata description} & \textbf{Please fill in this column} \\
\hline
C1 & Current code version & 0.0.2-archived \\
\hline
C2 & Permanent link to code/repository used for this code version & \url{https://github.com/ucl-bug/jwave/archive/refs/tags/0.0.2-archived.zip} \\
\hline
C4 & Legal Code License   & LGPL-3.0 \\
\hline
C5 & Code versioning system used & git \\
\hline
C6 & Software code languages, tools, and services used & Python 3 \\
\hline
C7 & Compilation requirements, operating environments \& dependencies & jax, plum-dispatch, jaxdf \\
\hline
C8 & If available Link to developer documentation/manual & \url{https://ucl-bug.github.io/jwave/} \\
\hline
C9 & Support email for questions & \url{a.stanziola@ucl.ac.uk}\\
\hline
\end{tabular}
\caption{Code metadata (mandatory)}

\end{table}

\newpage

}

\let\thefootnote\relax\footnotetext{$^*$\texttt{a.stanziola@ucl.ac.uk} }


\section{Motivation and significance}
\label{sec:motivation}

\subsection{Background}
The accurate simulation of wave phenomena has many interesting applications, from medical physics to seismology and electromagnetics, with the aim of either forecasting, for example, predicting an ultrasound field inside the brain \cite{aubry2022benchmark}, or performing parametric inference, for example, recovering material properties from acoustic measurements using full wave inversion  \cite{virieux2009overview}. Many numerical techniques for solving the wave equation have been developed over the years, including pseudospectral algorithms \cite{tabei2002k}, finite differences \cite{pinton2009heterogeneous, pichardo2017viscoelastic}, angular spectrum methods \cite{vyas2012ultrasound} and boundary element methods \cite{van2015fast}, to name a few.

Recently, there has been a  growing body of research at the intersection of numerical simulation and machine learning \cite{cranmer2020frontier, rackauckas2020universal, raissi2019physics}. The critical observation is that the machine learning community has developed many tools and techniques for high-dimensional inference. In particular, automatic differentiation, the class of algorithms often employed for neural network training and generally for automatic analytical gradient estimation, can be used to differentiate for any continuous parameter involved in a simulator \cite{innes2019differentiable, jax2018github}. This enables optimization or parameter identification of all simulator parameters, including the simulated field and other parameters that appear in the governing partial differential equation (PDE), as well as numerical parameters such as the finite difference stencil used to compute gradients. 

Simulators that allow for automatic differentiation can also be used inside a machine learning model. Examples include implementing implicit layers \cite{chen2018neural}, reinforcement learning \cite{lutter2021differentiable, murthy2020gradsim}, parameter identification \cite{heiden2021probabilistic}, inverse problems \cite{liang2019differentiable}, optimal control \cite{hu2019difftaichi}, construction of physics-based loss functions \cite{karpatne2017physics,hu2019difftaichi, holl2020learning}, and research into novel discretizations or neural network augmented simulators \cite{siahkoohi2019neural}.

\subsection{Aim}
Here we present j-Wave: a customizable Python simulator, written on top of the JAX library \cite{jax2018github} and the discretization framework JaxDF \cite{stanziola2021jaxdf}, for fast, parallelizable, and differentiable acoustic simulations. j-Wave solves both time-varying and time-harmonic forms of the wave equation with support for multiple discretizations, including finite differences and Fourier spectral methods. Custom discretizations, including those based on neural networks, can also be utilized via the JaxDF framework. The use of the JAX library gives direct support for  program transformations, such as automatic differentiation, Single-Program Multiple-Data (SPMD) parallelism, and just-in-time compilation. Lastly, since j-Wave is written in a language that follows the NumPy \cite{harris2020array} syntax, it is easy to adapt, enhance or re-implement any simulator stage.

\subsection{Related software}
There is a range of related software that can be used to simulate acoustic fields, and that can be used as an alternative or to complement j-Wave. In the Julia language, the SciML ecosystem has a variety of tools that can be used to construct differentiable acoustic simulators \cite{rackauckas2020universal}. In particular, the \texttt{ADSeismic.jl} \cite{zhu2021general} library focuses on seismic wave propagation and several inversion algorithms commonly used in the seismic field, and also includes the support for neural network representation of velocity models \cite{zhu2021integrating}. In Python, the Devito package \cite{lange2016devito} and the recently published Stride \cite{CUETO2022106855} library can be used to solve acoustic optimisation problems that scale over large super computing clusters, while SimPEG \cite{COCKETT2015142} can be used for geophysical parameter estimation. In JAX, several recent works have developed tools for simulation-based inference and differentiable simulations. These range from integrating JAX with FEniCS for finite elements simulations \cite{yashchuk2020bringing}, to differentiable molecular dynamics \cite{jaxmd2020} and fluid dynamics \cite{kochkov2021machine}.
\section{Software description}

\subsection{Governing equations}
\label{subsec:wave_eq}
j-Wave solves two different forms of the wave equation for time-varying and time-harmonic (i.e., single frequency) problems. For time-varying problems, j-Wave solves a linear system of coupled first-order PDEs that represent the conservation of mass and momentum, and a pressure density relation \cite{treeby2012modeling}:
\begin{align}
    \frac{\partial \mathbf{u}}{\partial t} &= - \frac{1}{\rho_0}\nabla p \\
    \frac{\partial \rho}{\partial t} &= -\rho_0\nabla \cdot \mathbf{u} + S_M  \\
    p &= c_0^2\rho \enspace.
\end{align}
Here $u$ is the acoustic particle velocity, $p$ is the acoustic pressure, and $\rho$ is the acoustic density. The acoustic medium is characterized by a spatially varying background density $\rho_0$ and sound speed $c_0$. The term $S_M$ represents a mass source field.

For time-harmonic simulations, j-Wave solves a form of the Helmholtz equation  constructed from the second-order wave equation including Stokes absorption:
\begin{equation}
    \frac{1}{c_0^2}\frac{\partial^2 p}{\partial t^2} = \nabla^2 p - \frac{1}{\rho_0}\nabla \rho_0 \cdot \nabla p + \frac{2\alpha_0}{c_0}\frac{\partial^3p}{\partial t^3} + \frac{\partial S_M}{\partial t}
\end{equation}
A time-harmonic solution is obtained by substituting $p = Pe^{-i\omega t}$, where $\omega$ is frequency in units of rad$\cdot$s$^{-1}$, giving
\begin{equation}
    -\frac{\omega^2}{c_0^2}P = \nabla^2 P - \frac{1}{\rho_0} \nabla \rho_0 \cdot \nabla P + \frac{2i\omega^3\alpha_0}{c_0} P - i \omega S_M.
\end{equation}
This equation accounts for acoustic absorption of the form $\alpha = \alpha_0 \omega^2$, where the absorption coefficient prefactor $\alpha_0$ has units of Np(rad/s)$^{-2}$m$^{-1}$.

\subsection{Numerical methods}

Solvers for the two governing equations given in Sec.\ \ref{subsec:wave_eq} are constructed using JaxDF \cite{stanziola2021jaxdf}. This is a discretization framework that decouples the mathematical definition of the problem from the underlying discretization. Currently, implementations of the differential operators are available for spectral and finite difference discretizations on a regular Cartesian grid. Alternatively, the user can provide a custom discretization compatible with the underlying operations required by the PDEs. That is, only linear discretizations are compatible with time-stepping and Krylov solvers, while non-linear discretizations can be used as physics informed models \cite{raissi2019physics, rackauckas2020universal}.

For time-varying problems, the wave equation is solved by integrating the first-order system of equations with a semi-implicit first-order Euler integrator. If a spectral or finite difference discretization is used, the fields are defined on a staggered grid to improve long-range accuracy \cite{treeby2010k} and avoid checker-board artifacts. Radiating boundary conditions are enforced by embedding the effect of a split-field perfectly matched layer (PML) on the time-stepping scheme \cite{tabei2002k}. When using a Fourier discretisation, j-Wave is equivalent to the implementation in the open-source k-Wave toolbox \cite{treeby2010k,treeby2012modeling}, including the use of a dispersion-corrected finite difference scheme for time integration. The user can further specify a generic measurement operator $f(u, \rho, p)$ to extract instantaneous values from the wavefield at each time step. 

For time-harmonic problems, if the underlying discretization of the Helmholtz operator is linear (for example, using Fourier or finite difference methods), the solver is a special case of linear inversion. In this case, j-Wave uses either GMRES or Bi-CGSTAB to compute the solution. These are matrix-free methods, meaning that the numerical matrix that represents the linear operator is never explicitly constructed. Again, radiating boundary conditions are imposed using a PML, by modifying the spatial gradients as in \cite{bermudez2007optimal}:
\begin{equation}
    \frac{\partial}{\partial x} \to \frac{1}{\gamma_x} \frac{\partial}{\partial x}
\end{equation}
where
\begin{equation}
    \gamma_x (x) = \begin{cases}
       1, & \text{if } |x| < a\\
       1 + \frac{i}{\omega}\sigma(x) & \text{if } a \leq |x|
    \end{cases},
\end{equation}
and $\sigma$ follows a power-law profile.

\subsection{JAX and automatic differentiation}

The fundamental idea of j-Wave is to provide a suite of differentiable, parallelizable and customizable acoustic simulators. These requirements are accomplished, in first instance, by writing the simulator in JAX \cite{jax2018github}, which provides a growing suite of tools for large-scale differentiable computations, including flexible AD, single-device parallelization, multi-device parallelization, and just-in-time compilation \cite{mpi4jax}. Furthermore, JAX can be considered an adaptable Python compiler that translates and transforms code. This allowed us to define a series of custom classes that can be overwritten or adapted by the user, while still being amendable of transformation.

All forward operators and simulation functions in j-Wave are differentiable through the use of JaxDF using both forward and backward automatic differentiation (AD). This allows the user to obtain gradients for any continuous parameter in the model. This includes both physical parameters, such as the acoustic pressure or sound speed, and numerical parameters, such as the stencils for finite differences or the filters used in Fourier methods. The gradient rules used for computation can also be freely customized.\footnote{Gradients obtained using reverse-mode AD have been shown to be equivalent to the ones obtained using the adjoint-state model \cite{zhu2021general}.}

Solving a linear system, such as the discretized Helmholtz equation, using an iterative solver is also beneficial for gradient calculation. JAX takes advantage of the implicit function theorem to differentiate through fixed-point algorithms with $O(1)$ memory requirements (that is, the intermediate steps of the iterative solver are not stored to compute the gradient). This is a major advantage when gradients of large-scale simulations are needed. See  \cite{blondel2021efficient} and references therein for a recent discussion of this topic.

\subsection{Software architecture}
\label{subsec:software_architecture}

The architecture of j-Wave can be divided into three main kinds of components: objects, operators, and solvers.

\begin{figure}[tp]
    \makebox[\textwidth][c]{\hspace*{-1.2cm}
        \scalebox{0.64}{\includegraphics[page=1]{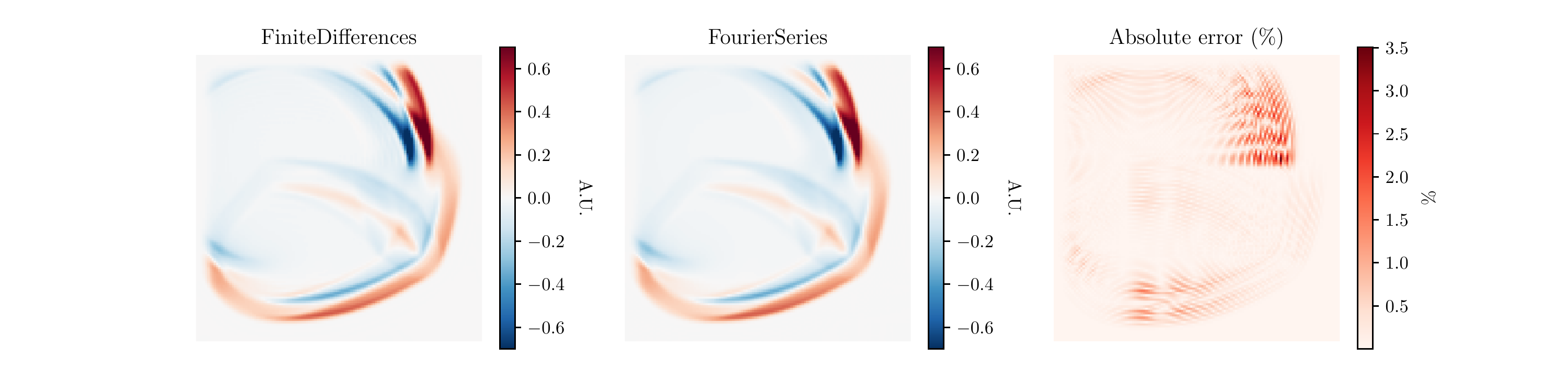}}
    }
    \caption{Comparison of the fields produced by j-Wave using 8th order accurate \texttt{FiniteDifferences} and \texttt{FourierSeries} representations on an initial value problem.
    \label{fig:fourier_vs_fd}
    }
\end{figure}

\begin{description}
\item[Objects: ]
Objects are variables that contain the numerical data that is used during the simulations. They are defined as classes registered to the JAX compiler as a custom pytree node. The primary objects are:
\begin{itemize}
  \item \texttt{Domain}: Defines a regular Cartesian grid with the specified grid spacing and number of points.
  \item \texttt{Medium}: Defines the \verb=sound_speed= and \verb=density= represented on the specified \verb=domain= along with the \verb=pml_size=.
  \item \texttt{Sources}: Defines the \verb=positions= and \verb=signals= for time varying mass sources within the specified \verb=domain=.
  \item \texttt{Sensors}: Defines the \verb=positions= of detectors placed on the grid.
  \item \texttt{TimeAxis}: Defines the time steps used for time-varying simulations.
\end{itemize}

Objects can be used as input variables to any JAX function and gradients can be taken with respect to their continuous parameters. They can be unpacked into their constituent numpy-like arrays using the \verb=jax.tree_util.tree_flatten= utility and constructed inside pure functions.

Some parameters are defined as \verb=Field= objects from JaxDF which define underlying discretizations. This includes \verb=medium.sound_speed= and the initial conditions \verb=p0= and \verb=u0=. The discretization used for the input objects governs the discretization used during the calculations. Currently, JaxDF supports \verb=FourierSeries=, \verb=FiniteDifferences= and \verb=Continuous= discretizations. However, it is straightforward to define custom field discretizations which are automatically compiled into their corresponding numerical implementations.

\item[Operators:] 
Operators are defined via JaxDF and implement a numerical algorithm that translates a symbolic operator into its corresponding numerical implementation, for a given type of input discretization. The implementation of the same operator for different discretizations is done using multiple-dispatch via \texttt{plum} \cite{wessel2022plum}, a programming technique that has been heavily popularized by C\#, Lisp and Julia \cite{Julia-2017}, using the \texttt{operator} decorator of JaxDF.

For example, a custom Laplacian operator for a 1D \texttt{FiniteDifferences} field can be implemented using type hints. 

\begin{lstlisting}[language=Python]
@operator
def laplacian(u: FiniteDifferences, params=[1, -2, 1]):
  k = params
  _u = u.on_grid
  _u = jnp.pad(_u, (1,1), 'constant', 0)
  v = k[0]*_u[:-2] + k[1]*_u[1:-1] + k[0]*_u[2:]
  return u.replace_params(v), params
\end{lstlisting}

Every function that uses the \texttt{laplacian} function will then utilize the custom user implementation if the input field is of the type \texttt{FiniteDifferences}.

\item[Solvers:]
There are two main solvers in j-Wave which solve the equations outlined in Sec.\ \ref{subsec:wave_eq}. These are also implemented as operators for convenience.

\begin{itemize}
  \item \texttt{simulate\_wave\_propagation}: Takes a \verb=medium= object (which internally defines the \verb=Domain=), along with \verb=Sources=, \verb=Sensors=, and \verb=TimeAxis= objects, and initial conditions \verb=p0= and \verb=u0= if non-zero, and computes the time varying acoustic field over the specified domain.
  \item \texttt{helmholtz\_solver}: Takes a \verb=medium= object (which internally defines the \verb=Domain=), \verb=source= field, and frequency \verb=omega= and computes the complex field over the specified domain. The \verb=source= field for the \texttt{helmholtz\_solver} is a \verb=Field= defined over the entire domain, and can be extracted from \verb=Sources= objects.
\end{itemize}

Simulations using these functions can be performed on CPU, GPUs, and TPUs, with efficient just-in-time compilation, natively compatible with the JAX ecosystem. The functions are also amendable to same-device or multiple-devices parallelization, via the JAX decorators \texttt{vmap} and \texttt{pmap} \cite{jax2018github}. Check-pointing can also be applied at each step to reduce the memory requirements for back-propagation.

\end{description}

\subsection{Accuracy}

The accuracy of the pseudo-spectral and finite difference solvers has been evaluated both for time-varying and for time-harmonic problems. In the first case, the pseudo-spectral  numerical solver is equivalent to k-Wave \cite{treeby2010k,treeby2012modeling} and numerical simulations agree to machine precision. When finite difference methods are employed, the simulation error is dependent on many factors, other than the implementation itself, such as number of grid points per wavelength, the finite difference coefficients, etc. An illustrative comparison of the wavefields produced for an initial value problem in a medium with a heterogeneous sound speed is shown in Fig. \ref{fig:fourier_vs_fd}.

\begin{figure}[tp]
    \makebox[\textwidth][c]{\hspace*{-1.2cm}
        \scalebox{0.64}{\includegraphics[page=2]{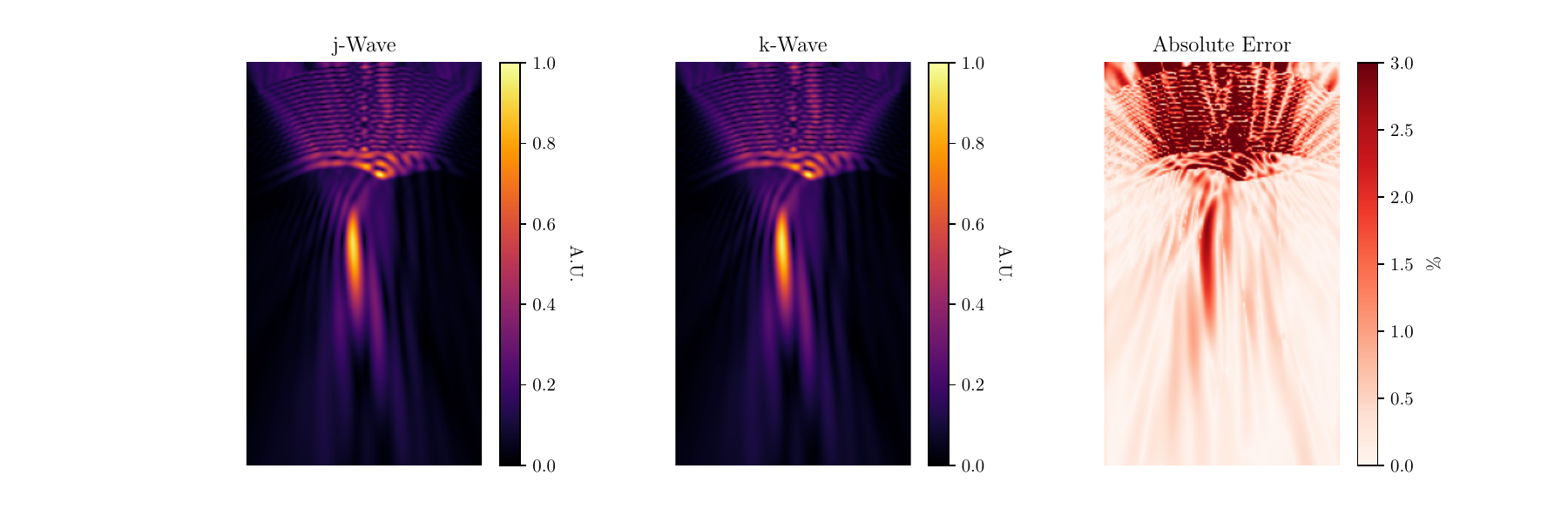}}
    }
    \caption{Comparison of the field amplitudes predicted by j-Wave and k-Wave for a focused transducer after propagation through an aberrating skull layer. Adapted from \cite{aubry2022benchmark}.
    \label{fig:skull_helmholtz_comparison}
    }
\end{figure}

For the Helmholtz equation, a comprehensive comparison of j-Wave against other wave models (including k-Wave) was conducted as part of the inter-comparison effort described in \cite{aubry2022benchmark}. For homogeneous material properties, the maximum difference against k-Wave is typically much less than $1\%$. For heterogeneous properties, the difference depends on which parameters are heterogeneous and the strength of the heterogeneity. Differences are slightly larger for a heterogeneous density (compared to heterogeneous sound speed or absorption). This is likely due to the different way the ambient density term is treated and evaluated on a staggered grid between the two softwares. A representative example showing results for a 3D simulation using j-Wave and k-Wave is given in Fig. \ref{fig:skull_helmholtz_comparison}. This example includes a bone layer   with an incident field produced by a focused transducer driven at 500kHz (Benchmark 7 of \cite{aubry2022benchmark}). In this case, the difference between the two simulations inside the brain is within 3\%.

\section{Illustrative Examples}
\label{sec:examples}

\subsection{Initial value problems and image reconstruction using time reversal}
\label{subsec:code_snippets}

To demonstrate the process of defining and running a simulation using j-Wave, we start with a simple initial value problem in a homogeneous medium as encountered in, e.g., photoacoustics \cite{cox2005fast}. Similarly to k-Wave \cite{treeby2010k}, j-Wave requires the user to specify a computational domain where the simulation takes place. This is done using the \texttt{Domain} data class inherited from JaxDF as shown in Listing \ref{lst:domain}. The inputs for the constructor are the size of the domain in grid points in each spatial direction and the corresponding discretization step.

\begin{minipage}{\linewidth}
\begin{lstlisting}[language=Python,caption={Defining the simulation domain.},label={lst:domain}]
from jwave.geometry import Domain

N, dx = (128, 128), (0.1e-3, 0.1e-3)
domain = Domain(N, dx)
\end{lstlisting}
\end{minipage}

The next step is to define the medium properties. This is done using the \texttt{Medium} class as shown in Listing \ref{lst:medium}.
\begin{minipage}{\linewidth}
\begin{lstlisting}[language=Python,caption={Defining the medium properties.},label={lst:medium}]
from jwave.geometry import Medium

medium = Medium(domain=domain, sound_speed=1500.0)
\end{lstlisting}
\end{minipage}

For time-varying problems, a \texttt{TimeAxis} object also needs to be defined, which sets the time steps used in the time-stepping scheme of the numerical simulation. This object can be constructed from the medium for a given Courant-Friedrichs-Lewy (CFL) number as shown in Listing \ref{lst:time_axis} to ensure that the time-stepping scheme is stable

\begin{minipage}{\linewidth}
\begin{lstlisting}[language=Python,caption={Defining the time axis.},label={lst:time_axis}]
from jwave.geometry import TimeAxis

time_axis = TimeAxis.from_medium(medium, cfl=0.3)
\end{lstlisting}
\end{minipage}

The next optional step is to define a \texttt{Sensors} object. This is done using the \texttt{Sensors} class as shown in Listing \ref{lst:sensors}, which defines the grid points within the domain where the field values are returned (custom sensor definitions can also be used). If no sensors are defined, the code returns a \verb=Field= for each time-step.

\begin{minipage}{\linewidth}
\begin{lstlisting}[language=Python,caption={Defining sensors.},label={lst:sensors}]
from jwave.geometry import _points_on_circle, Sensors

num_sensors, radius, center = 32, 40, (64, 64)
x, y = _points_on_circle(num_sensors, radius, center)
sensors = Sensors(positions=(jnp.array(x), jnp.array(y)))
\end{lstlisting}
\end{minipage}

Finally, the initial pressure distribution must be defined. This is done by populating a \texttt{jax.numpy.ndarray} the same size as the domain, and then passing this to the appropriate discretization. In Listing \ref{lst:p0}, the initial pressure is set to  the weighted sum of four binary disks and defined as a \texttt{FourierSeries} field. The field information is used when calling operators to choose the correct numerical implementations. The simulation setup is depicted in Fig. \ref{fig:code_snippets_figures} (a).

\begin{minipage}{\linewidth}
\begin{lstlisting}[language=Python,caption={Defining the initial pressure distribution as a Fourier series Field.},label={lst:p0}]
from jwave.geometry import _circ_mask
from jwave import FourierSeries

mask1 = _circ_mask(N, 8, (50,50))
mask2 = _circ_mask(N, 5, (80,60))
mask3 = _circ_mask(N, 10, (64,64))
mask4 = _circ_mask(N, 30, (64,64))
p0 = 5.*mask1 + 3.*mask2 + 4.*mask3 + 0.5*mask4
p0 = FourierSeries(p0, domain)
\end{lstlisting}
\end{minipage}

To run the simulation, the solver \texttt{simulate\_wave\_propagation} is called with the appropriate inputs as shown in Listing \ref{lst:jit_and_run}. Here, a wrapper is defined around it, to highlight how to create arbitrary callables that are just-in-time compiled using \texttt{jax.jit}. The recorded acoustic signals are shown in Fig. \ref{fig:code_snippets_figures} (b).

\begin{minipage}{\linewidth}
\begin{lstlisting}[language=Python,caption={Just-in-time compiling and running the simulation.},label={lst:jit_and_run}]
from jwave.acoustics import simulate_wave_propagation

@jit
def compiled_simulator(medium, p0):
  return simulate_wave_propagation(
    medium, time_axis, p0=p0, sensors=sensors)
    
sensors_data = compiled_simulator(medium, p0)
\end{lstlisting}
\end{minipage}

\subsection{Automatic differentiation}
As mentioned, gradients can be evaluated with respect to any input parameters: all that is needed is to define a scalar loss function. In Listing \ref{lst:time_reversal}, the use of the wave equation adjoint as a simple imaging algorithm for the forward problem defined in Sec.\ \ref{subsec:code_snippets} is demonstrated following the discretize-then-optimize approach \cite{betts2005discretize, rackauckas2020universal}. Note that the user can always define a custom adjoint function for the forward operator if required.

Gradients for the initial pressure alone can be easily computed by wrapping a new function around the simulator and using the \texttt{jax.grad} decorator. In this example, noise is added to the data before inverting the model.

\begin{lstlisting}[language=Python,caption={Use of the adjoint model as a simple imaging algorithm.},label={lst:time_reversal}]
def solver(p0):
  return simulate_wave_propagation(
    medium, time_axis, p0=p0, sensors=sensors)

@jit  # Compile the whole algorithm
def lazy_imaging_algorithm(measurements):
  # Mask out elements outside the sensors ring
  mask = _circ_mask(N, 39, (64, 64))
  mask = np.expand_dims(mask, -1)

  def mse_loss(p0, measurements):
    p0 = p0.replace_params(p0.params * mask)
    p_pred = solver(p0)
    return 0.5 * jnp.sum((p_pred - measurements)**2)
    
  # Start from an empty field
  p0 = FourierSeries.empty(domain)
  # Take the gradient
  p_grad = grad(mse_loss)(p0, measurements)
  return -p_grad

# Reconstruct initial pressure distribution
recon_image = lazy_time_reversal(noisy_data)
\end{lstlisting}

\begin{figure*}[t]
    \makebox[\textwidth][c]{\hspace*{-1.2cm}
        \scalebox{0.64}{\includegraphics[page=3]{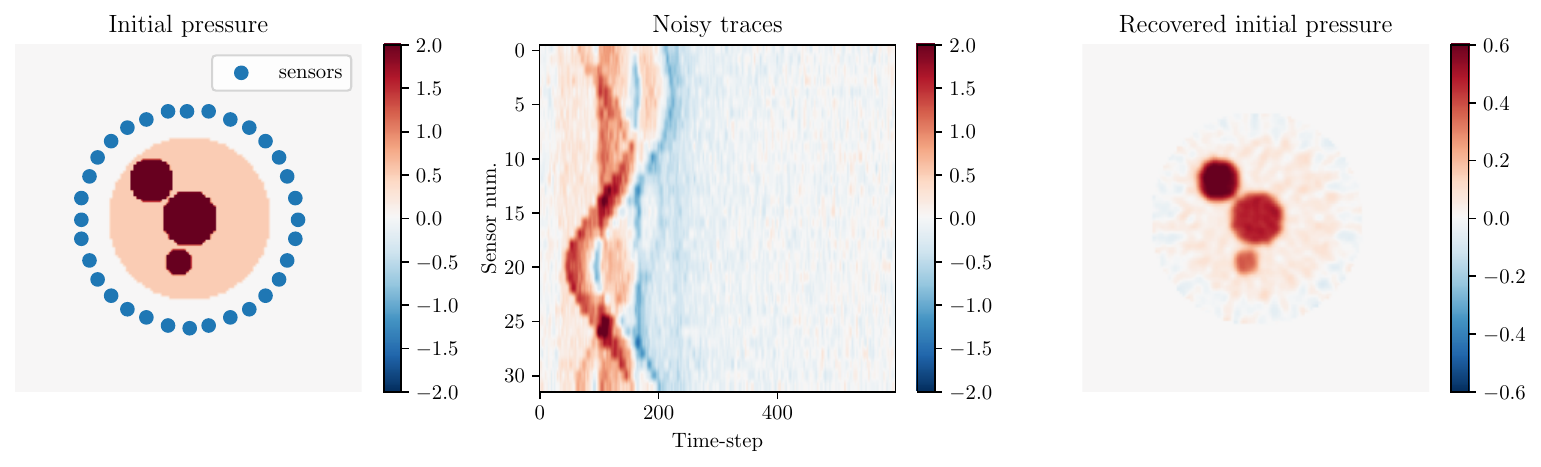}}
    }
 \caption{\label{fig:code_snippets_figures} Example of workflow to simulate an initial value problem and invert it using automatic differentiation. From left to right: Simulation setup; Recorded acoustic signals with additive colored noise; Reconstructed initial pressure distribution from noisy data.}
\end{figure*}

The reconstructed initial pressure is shown in Fig. \ref{fig:code_snippets_figures} (c).

\subsection{Prototyping full-wave inversion algorithms}
\begin{minipage}{\linewidth}
\begin{lstlisting}[language=Python,caption={Defining an objective function for full-wave inversion.},label={lst:fwi_obj_func}]
from jwave.signal_processing import analytic_signal

def loss_func(params, source_num):
  # This contains the simulator function
  p = single_source_simulation(get_sound_speed(params), source_num)

  # Get envelopes of data and simulated signals
  p = jnp.abs(analytic_signal(p, 0))
  pred = jnp.abs(analytic_signal(p_data[source_num], 0))

  # MSE on envelopes
  return jnp.sum(jnp.abs(p - pred)**2) 
    
loss_with_grad = jax.value_and_grad(loss_func)
\end{lstlisting}
\end{minipage}
One of the most exciting features of j-Wave is its (almost) total differentiability. Besides applications in machine learning, differentiability means that full waveform inversion methods can be easily prototyped. For example, to mitigate cycle skipping it has been proposed to use an $\ell_2$ loss on the modulus of the complex analytic signal associated with the data residual \cite{bedrosian1962analytic, chi2014full}. This can be implemented by defining an appropriate objective function as shown in Listing \ref{lst:fwi_obj_func}.

\begin{minipage}{\linewidth}
\begin{lstlisting}[language=Python,caption={Gradient descent using AD.},label={lst:fwi_grad_desc}]
@jax.jit
def update(opt_state, key, k):
  v = get_params(opt_state)
  src_num = random.choice(key, num_sources)

  loss_with_grad = jax.value_and_grad(loss_func, argnums=0)
  lossval, gradient = loss_with_grad(v, src_num)

  gradient = smooth_fun(gradient)
  return lossval, update_fun(k, gradient, opt_state)
\end{lstlisting}
\end{minipage}

Because it is possible to differentiate through arbitrary computations, evaluating the gradient of this expression is done using backward-mode AD. Low-pass filtering of the FWI speed of sound gradients can also be used to improve the convergence towards the true speed of sound distribution \cite{alkhalifah2014scattering}. Again, we can seamlessly include smoothing of the gradients in the update function that is run at each iteration of gradient descent as shown in Listing \ref{lst:fwi_grad_desc}. 

\begin{figure}[t!]
    \centering
    \makebox[\textwidth][c]{\hspace*{-1.2cm}
        \scalebox{0.75}{\includegraphics[page=4]{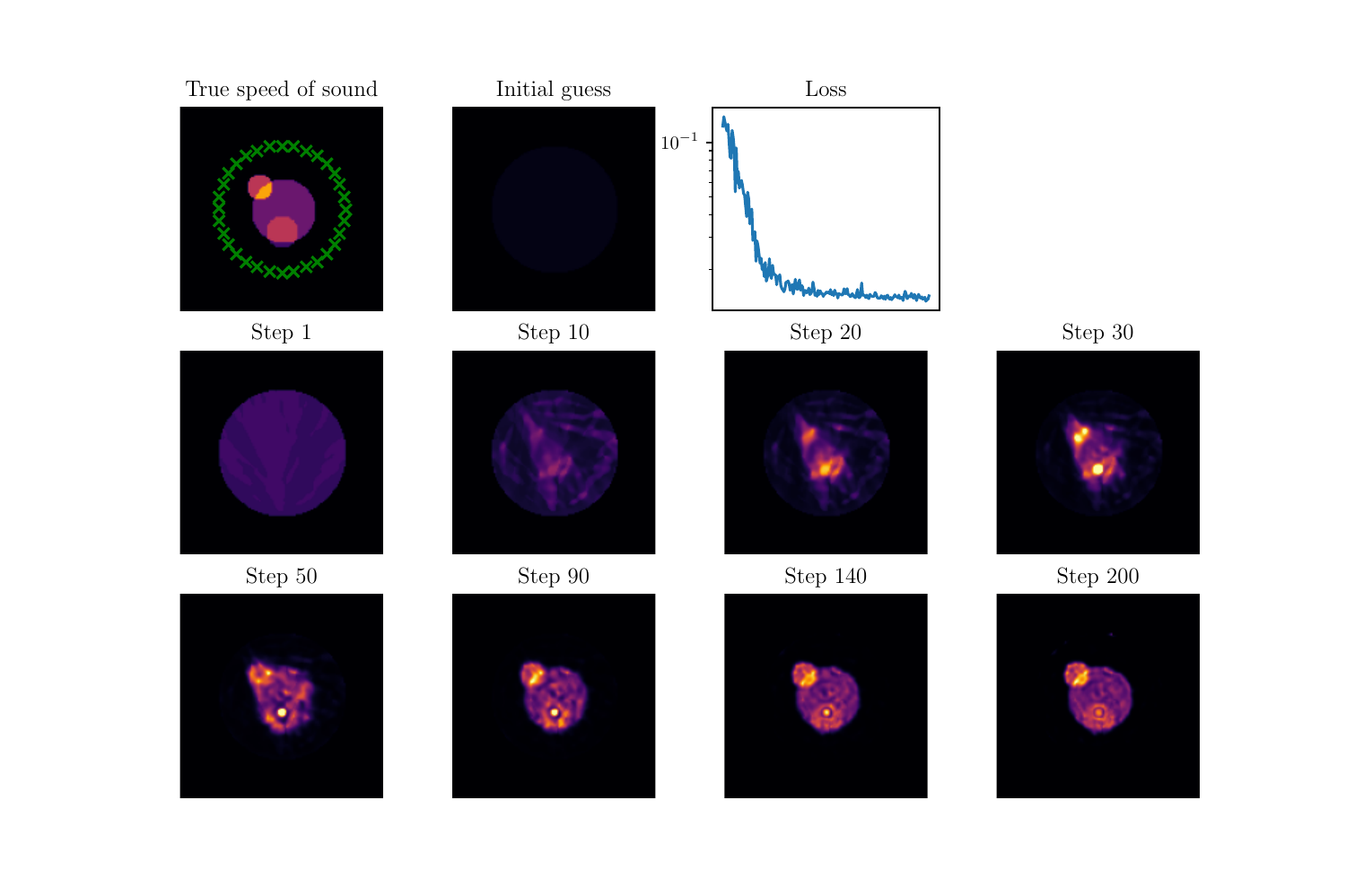}}
    }
    \caption{Full wave inversion using an envelope-based objective function and speed of sound gradient smoothing.}
    \label{fig:fwi_example}
\end{figure}

The results of this FWI algorithm on a noisy synthetic dataset are given in Fig. \ref{fig:fwi_example}. Note that this example is only intended to highlight the ability to take gradients of arbitrary computations using a discretize-then-optimize approach.

\subsection{Focusing of time-harmonic simulations}

As a final example, we demonstrate the differentiability of the time-harmonic solver. We transmit waves from a set of $n$ transducers, that act as monopole sources: that means that we can define a complex weighting vector, that defines the amplitude and phase of the sources
\begin{equation}
    \mathbf a = (a_0, \dots, a_n), \qquad a_i \in \mathbb{C}, \; \|a_i\| < 1
\end{equation}
such that $\rho(\mathbf a)$ is the transmitted wavefield. The unit norm constraint is needed to enforce the fact that each transducer has an upper limit on the maximum power it can transmit. One could use several methods to represent this vector and its constraint. Here, we use the following parameterization:
\begin{equation}
    a_j(\rho_j, \theta_j) = \frac{e^{i\theta_j}}{1 + \rho_j^2},
\end{equation}
where $\rho_j$ and $\theta_h$ are real variables. 

\begin{figure*}[t]
    \centering
    \makebox[\textwidth][c]{\hspace*{-0cm}
        \scalebox{.95}{\includegraphics[page=5]{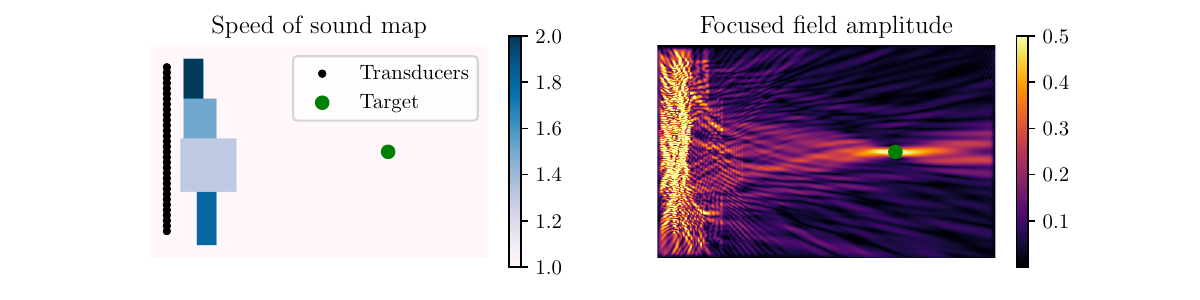}}
    }
 \caption{\label{fig:helmoltz} Example where the differentiability of the time-harmonic simulator is used. (a) Simulation setup, with a line of point transducers, heterogeneous sound speed and a focusing target; (b) The amplitude of the acoustic field after optimizing the transmit apodization.}
\end{figure*}

Often, one wants to find an apodization vector which returns a field having certain properties. For example, in transcranial neurostimulation one may want to maximize the acoustic power delivered to a certain location: this is the setup that we'll use in this example (see Fig. \ref{fig:helmoltz} (a) ).

Let's call $\mathbf p\in\mathbb{R}^2$ the point where we want to maximize the wavefield. For a field $\phi(\mathbf x,\mathbf a)$ generated by the apodization $\mathbf a$, the optimal apodization is then given by
\begin{equation}
    \hat {\mathbf a} = \operatorname*{arg\,max}_{\mathbf a} \|\phi(\mathbf p, \mathbf a) \|.
\end{equation}
This defines the loss function that we are going to minimize using gradient descent. The full code for this example is given in the notebook \texttt{helmholtz\_solver\_differentiable.ipynb}, in the examples folder. The resulting wavefield after the optimization is shown in Fig. \ref{fig:helmoltz} (b).

\section{Impact}
\label{sec:impact}

j-Wave combines several ideas from the machine learning and inverse problems communities, and can be used to investigate numerical and physical problems revolving around acoustic phenomena. The software is open-source and is based on JAX, which uses an interface that closely follows the widely used NumPy package \cite{harris2020array}. This means that interested researchers can customize the software to their needs using a familiar syntax.

As a forward solver, j-Wave  can be used as a simple pseudo-spectral acoustic simulator to perform numerical acoustic experiments. The software can simulate wave propagation in homogeneous and heterogeneous media, both in the frequency domain and in the time domain.

The differentiability of the solver can be exploited for a variety of tasks. By taking gradients with respect to the acoustic parameters, j-Wave can perform discrete sensitivity analyses or can be used to learn machine-learning models that perform model-based image inversion. Similarly, gradients with respect to the source parameters can be used for model-based optimal control and training reinforcement learning agents that interact with an acoustic setup. 

j-Wave as a differentiable forward model can also be exploited for uncertainty quantification. Besides Monte Carlo methods that can be accelerated in j-Wave using single-device and multiple-device parallel transformations, there is a growing body of techniques that are being developed to exploit simulation gradients for simulation-based inference \cite{cranmer2020frontier, gerlach2020koopman}. For example, in \cite{giordano2016uncertainty}, the use of linear uncertainty propagation (LUP) was proposed as a meta-programming method to endow arbitrary (differential) simulations with uncertainty propagation in the Julia language \cite{bezanson2017julia}. Supporting forward automatic differentiation allows LUP to be implemented  with minimal memory requirements for simulations that depend on a small number of parameters (e.g., uncertainty on the background speed of sound).

Since the operators relevant for acoustic simulations are implemented with JaxDF \cite{stanziola2021jaxdf}, it is possible to experiment with arbitrary discretizations that contain tunable parameters. This could be leveraged for a variety of tasks such as reduction of memory requirements, computational acceleration, or parameter inference. This is further aided by the possibility of overriding the behaviour of operators for existing or user-defined discretizations. For example, a similar approach has been used recently in computational fluid dynamics, where the authors trained a neural network-based adaptive finite-difference scheme to perform accurate simulations on coarser collocation grids \cite{kochkov2021machine}. Alternatively, one could employ learned error-correction schemes \cite{siahkoohi2019neural}, directly optimize the 
stencils of a finite difference scheme \cite{jo1996optimal}, or learn a preconditioner for the discretized Helmholtz equation \cite{azulay2021multigrid}.

Operators that represent a PDE, such as the Helmholtz operator, can also be constructed for arbitrary nonlinear discretizations, allowing the application of Physics Informed Neural Networks to solve the acoustic problem \cite{raissi2019physics}.

\section{Conclusions}
\label{sec:conclusions}

An open-source differentiable acoustic simulator called j-Wave is presented that solves both time-harmonic and time-varying forms of the wave equation. The simulator is written in JAX and is compatible with machine learning libraries. Furthermore, it provides a differentiable implementation of the time-harmonic acoustic operator (Helmholtz operator) that can be used either with both linear and non-linear arbitrary discretizations, including ones depending on a set of tunable parameters. We expect j-Wave to be a useful tool for a wide range of acoustic-related lines of research: from the investigation of numerical algorithms and machine learning ideas, to the design of acoustic imaging techniques and materials.


\section{Conflict of Interest}
We wish to confirm that there are no known conflicts of interest associated with this publication and there has been no significant financial support for this work that could have influenced its outcome.

\section*{Acknowledgements}

This work was supported by the Engineering and Physical Sciences Research Council (EPSRC), UK, grant numbers EP/S026371/1 and EP/T022280/1.




\bibliographystyle{elsarticle-num} 
{\footnotesize \bibliography{biblio.bib}}

\newpage

\end{document}